\documentclass[12pt,preprint]{emulateapj}
\usepackage{graphicx}

\shorttitle{Phase Variations of 55 Cnc}
\shortauthors{Stephen R. Kane et al.}
\slugcomment{Submitted for publication in the Astrophysical Journal}

\begin{document}

\title{Planetary Phase Variations of the 55 Cancri System}
\author{
  Stephen R. Kane\altaffilmark{1},
  Dawn M. Gelino\altaffilmark{1},
  David R. Ciardi\altaffilmark{1},
  Diana Dragomir\altaffilmark{1,2},
  Kaspar von Braun\altaffilmark{1}
}
\email{skane@ipac.caltech.edu}
\altaffiltext{1}{NASA Exoplanet Science Institute, Caltech, MS 100-22,
  770 South Wilson Avenue, Pasadena, CA 91125}
\altaffiltext{2}{Department of Physics \& Astronomy, University of
  British Columbia, Vancouver, BC V6T1Z1, Canada}


\begin{abstract}

Characterization of the composition, surface properties, and
atmospheric conditions of exoplanets is a rapidly progressing field as
the data to study such aspects become more accessible. Bright targets,
such as the multi-planet 55~Cancri system, allow an opportunity to
achieve high signal-to-noise for the detection of photometric phase
variations to constrain the planetary albedos. The recent discovery
that that inner-most planet, 55 Cancri e, transits the host star
introduces new prospects for studying this system. Here we calculate
photometric phase curves at optical wavelengths for the system with
varying assumptions for the surface and atmospheric properties of 55
Cancri e. We show that the large differences in geometric albedo
allows one to distinguish between various surface models, that the
scattering phase function cannot be constrained with foreseeable data,
and that planet b will contribute significantly to the phase variation
depending upon the surface of planet e. We discuss detection limits
and how these models may be used with future instrumentation to
further characterize these planets and distinguish between various
assumptions regarding surface conditions.

\end{abstract}



\section{Introduction}

The discovery of transiting exoplanets was a breakthrough in the
exoplanet field. This is not simply because it provided an additional
avenue through which to secure their detection, but because of the
vast amount of planetary characterization information they yield
access to. The investment in improving photometric techniques and
precision for transit surveys has led to the peripheral consequence of
detecting more subtle planetary signatures. Photometric phase curves
of exoplanets is one such signature, though technically difficult to
achieve due to the relatively low signal amplitudes. Examples of
observed phase variations in the Infra-Red (IR) from Spitzer
observations include HD~189733b \citep{knu09a} and HD~149026b
\citep{knu09b}. Examples in the optical include Kepler observations of
HAT-P-7b \citep{wel10} and phase variations detected in the light
curve of CoRoT-1b \citep{sne09}.

The most valuable transiting planets are the ones which orbit bright
host stars since they lend themselves towards greater signal-to-noise
measurements. Follow-up of known radial velocity (RV) planets is a key
way to achieve this goal \citep{kan09}. A recent example is that of
the planets orbiting 55~Cancri (HD~75732, HIP~43587, HR~3522,
hereafter 55~Cnc), a bright ($V = 5.95$) G8 dwarf star.  \citet{val05}
predict a relatively old age for the 55~Cnc system of $9.5 \pm
4.4$~Gyr, although \citet{fis08} estimate a more modest age range of
2--8~Gyr. Thus this is an interesting system since the host is of
similar spectral type and age to our own sun and yet the planetary
configuration is substantially different from our own system.

The first planet in the system was discovered by \citet{but97} and the
second and third were detected by \citet{mar02}. The fourth
Neptune-mass planet, 55~Cnc~e, was found by \cite{mca04}, with an
originally deduced orbital period of 2.82~days. The fifth planet was
discovered by \citet{fis08}, leaving us with the currently known
five-planet system. Subsequent study of the aliases in the power
spectrum of the RV data by \citet{daw10} found that the true period of
the e planet most likely 0.74~days rather than 2.82~days. This was
confirmed by \citet{win11} who successfully detected the transit of
55~Cnc~e using high-precision photometry from the Microvariability and
Oscillations of STars (MOST) satellite \citep{wal03}. Further
confirmation appeared through detection of the transit in the IR via
Spitzer observations \citep{dem11b}.

This system presents an opportunity to search for phase variations for
a known system around a very bright star where we know much more about
the relative planetary inclinations in the system than other similar
multi-planet systems. Phase variations for gas giants have been
described by \citet{kan10} and \citet{kan11} and the albedos and heat
redistribution properties for such planets are discussed by
\citet{cow11}. Here we apply these techniques to model the phase
variations of the 55~Cnc system using the revised orbital parameters.
\citet{win11} report the detection of phase modulations which are
inconsistent with their expectations, though more phase coverage to
improve the amplitude measurement and account for the presence of the
outer planets would assist in their interpretation. The flux ratio of
a planet with radius $R_p$ to the host star measured at wavelength
$\lambda$ is defined as
\begin{equation}
  \epsilon(\alpha,\lambda) \equiv
  \frac{f_p(\alpha,\lambda)}{f_\star(\lambda)}
  = A_g(\lambda) g(\alpha,\lambda) \frac{R_p^2}{r^2}
  \label{fluxratio}
\end{equation}
where $\alpha$ is the phase angle of the planet. This flux ratio
consists of three major components; the geometric albedo
$A_g(\lambda)$, the phase function $g(\alpha,\lambda)$, and the
inverse-square relation to the star--planet separation $r$. In
Section \ref{system} we describe the system characteristics which are
used as input into the flux ratio model for each of the planets. In
Section \ref{phase} we calculate the total system phase variations
based upon three different models of the inner planet. Finally, in
Section \ref{signal}, we assess the detectability of the individual
and combined planetary signals and future prospects for discriminating
between the surface models for the e planet.


\section{Science Motivation}
\label{motivation}

Detection of planetary phase variations are currently a challenge to
detect due to their relatively low amplitude. It is reasonable
therefore to assess the science return from such difficult
observations. For planets which undergo secondary eclipses, the albedo
may be determined through careful modeling of the eclipse data (see
for example \citet{dem11a}). This has the distinctive advantage of
being able to measure the key planetary parameters of radius and
orbital inclination. However, without complete phase variation
analysis, this albedo determination is generally only valid at or near
the sub-stellar point of the planet. Albedos are sensitive to such
aspects as cloud formation depth, reflective condensates in the upper
atmosphere, and the scattering properties of the
surface/atmosphere. Thus there is a strong degeneracy between the
albedo and back-scattering properties when the planet is only observed
at zero phase angle. By determining the phase function through
precision observations, one may gain further insight into these
aspects which are inaccessible via secondary eclipse observations.

Furthermore, continued observations of the phase function contain
information on the combined reflective properties for all planets in
the system. For planets whose presence and/or orbital properties are
unknown, this can lead to ambiguity in the interpretation of the data
\citep{kan10}. However, in cases such as the 55 Cnc system, the
planets presence and orbital parameters are well determined and thus
the degeneracy can be removed leading to constraints on the scattering
properties or the outer planets. As we shall demonstrate, planet b
contributes significantly to the total phase variation in this case.
Even if the planets are not known to transit, the phase curves can be
used to constrain the inclination of the orbits \citep{kan11}.


\section{System Characteristics}
\label{system}

Here we describe the system characteristics which are used as input
for the phase model, both measured and derived. These are summarized
in Table \ref{planchar}, along with the predicted phase variation
amplitudes which are described in more detail in Section \ref{phase}.

\begin{deluxetable*}{ccccccccccccc}
  \tablecolumns{12}
  \tablewidth{0pc}
  \tablecaption{\label{planchar} Planetary Orbital Parameters and
    Derived Characteristics}
  \tablehead{
    \colhead{Planet} &
    \colhead{$P\,^{\dagger}$} &
    \colhead{$M_p \sin i\,^{\dagger}$} &
    \colhead{$a\,^{\dagger}$} &
    \colhead{$e\,^{\dagger}$} &
    \colhead{$\omega\,^{\dagger}$} &
    \colhead{$R_p\,^{\ddagger}$} &
    \colhead{$i$} &
    \colhead{$A_g\,^{*}$} &
    \multicolumn{3}{c}{Flux Ratio ($10^{-6}$)} \\
    \colhead{} &
    \colhead{(days)} &
    \colhead{($M_J$)} &
    \colhead{(AU)} &
    \colhead{} &
    \colhead{(deg)} &
    \colhead{($R_J$)} &
    \colhead{(deg)} &
    \colhead{} &
    \colhead{Rocky} &
    \colhead{Molten} &
    \colhead{Atmosphere}
  }
  \startdata
  e & 0.736537 & 0.026 & 0.016 & 0.17  & 181 & 0.18 & 90.0 & 0.15 & 3.37 & 20.23 & 4.89 \\
  b & 14.6507  & 0.825 & 0.115 & 0.010 & 139 & 1.10 & 87.5 & 0.16 & --   & --    & 3.26 \\
  c & 44.364   & 0.171 & 0.240 & 0.005 & 252 & 1.01 & 88.8 & 0.17 & --   & --    & 0.70 \\
  f & 259.8    & 0.155 & 0.781 & 0.30  & 180 & 0.96 & 89.6 & 0.22 & --   & --    & 0.11 \\
  d & 5169.0   & 3.82  & 5.74  & 0.014 & 186 & 1.07 & 89.9 & 0.50 & --   & --    & 0.004
  \enddata
  \tablenotetext{$\dagger$}{From Table 10 of \citet{daw10}.}
  \tablenotetext{$\ddagger$}{With the exception of planet e, based upon
    \citet{bod03} models.}
  \tablenotetext{*}{Mean geometric albedo assuming a thick atmosphere.}
\end{deluxetable*}


\subsection{Orbital Considerations}

We adopt the complete system orbital solution shown in Table 10 of
\citet{daw10}. This is a self-consistent model which converges on the
solution which includes the correct period for planet e of
0.74~days. The fit also has a smaller rms scatter of the residuals
than the equivalent fit which reaches the old e period of 2.82~days.
These orbital properties are shown in Table \ref{planchar}, including
the period $P$, the mimimum mass $M_p \sin i$, the semi-major axis
$a$, the eccentricity $e$, and the argument of periastron
$\omega$. Note that the phase models presented here also account for
the eccentricities present in the orbits. For the host star, we adopt
a stellar mass of $M_\star = 0.94 \pm 0.05 \ M_\odot$ \citep{fis08}
and a stellar radius of $R_\star = 0.943 \pm 0.010 \ R_\odot$
\citep{von11}.


\subsection{Planetary Radii}

An important property for considering the amplitude of the
planet-to-star flux ratio from a given planet is the planetary
radius. This quantity is normally only available for a planet whose
transits reveal it to us, but can also be derived from the estimated
planetary mass and stellar properties for the planet in question. For
planet e, we adopt the radius measured by \citet{win11} of
2.00~$R_\oplus$, or 0.179~$R_J$. We select this radius rather than
that measured by \citet{dem11b} because we are considering phase
variation effects at similar passbands to MOST rather than
Spitzer. For the more massive outer planets in the system, we
calculate radii estimates based upon the models of \citet{bod03} which
take into account both the planetary mass and the stellar flux
received at their respective semi-major axis. The results of these
calculations are shown in Table \ref{planchar}.


\subsection{Orbital Inclinations}

The orbital inclinations of the planets are unknown except for that of
planet e. We adopt the value of $i = 90\degr$ for this planet from
\citet{win11}. The evidence thus far is that none of the outer planets
transit the host star. Here we assume that this is indeed the
case and calculate the maximum inclination which satisifies this
criteria using the methods described in \citet{kan08}. The inclination
in this case is then given by
\begin{equation}
  \cos i = \frac{R_p + R_\star}{r}
\end{equation}
where $r$ is the star--planet separation, given by
\begin{equation}
  r = \frac{a (1 - e^2)}{1 + e \cos f}
\end{equation}
and is evaluated at $\omega + f = \pi / 2$.

The above assumption for the maximum inclination hinges somewhat on
the system orbits being close to coplanar. Astrometry performed by
\citet{mca04} indicate that planet d may be misaligned with the
edge-on orbit of planet e, although this is based upon preliminary
work and the outer-most planet contributes a negligible amount of flux
to the combined phase curve. The results from the Kepler mission have
revealed many multiple-transiting system which implies that those
systems are remarkably coplanar \citep{lat11}. An excellent example is
the six-planet system orbiting Kepler-11 \citep{lis11}. Theoretical
modeling performed by \citet{tre11} independently supports the claim
by \citet{lis11} that near-zero mutual inclinations are favored for
multi-planet systems. Stability analyses of RV systems have also been
used to determine coplanarity, such as the cases of the HD~10180
system \citep{lov11} and GJ~876 system \citep{bea09}. It should be
noted however that dynamical stability over long timescales can be
achieved through interaction of giant planets or the influence of an
external perturber \citep{gui11,mal11}.


\subsection{Geometric Albedos}

Planetary albedos can span a very large range of values depending upon
both the surface conditions and location of the planet. The
theoretical models of \citet{sud05} show that there is a dependence of
gas giant geometric albedos on star--planet separation due to the
removal of reflective condensates from the upper atmosphere. This was
quantified by \citet{kan10} who also generalized this dependence to
eccentric orbits. Note that this does not take into account a variable
surface albedo or the thermal response of the surface/atmosphere to
changing incident flux. We use these models to estimate the geometric
albedos for the four outer-most planets.

However, the super-Earth planet e is a special case due to its smaller
size and proximity to the host star. In Section \ref{phase} we
consider three possible surface models which entail different albedos
and scattering properties. The models are that of a rocky surface, a
molten suface, and a thick atmosphere. The atmosphere model uses the
geometric albedos described above with the same star--planet
separation dependence. Shown in Table \ref{planchar} are the mean
calculated albedos for all of the planets based upon the thick
atmosphere assumption. The rocky and molten surface models use the
measured albedos of Mercury and Io as templates respectively, where
the values were extracted from the JPL HORIZONS System\footnote{\tt
  http://ssd.jpl.nasa.gov/?horizons}.  Note that the composition and
tidal forces on Io lead to a highly variable surface albedo
\citep{sim01}. However, we are considering the integrated flux from
the planet and so the mean albedo of Io is a useful approximation.
Mercury's geometric albedo is 0.106 and has a density of
5.427~g\,cm$^{-3}$. For Io, the geometric albedo is 0.6 and the
density is 3.530~g\,cm$^{-3}$. The density for 55~Cnc~e is calculated
from the properties in Table~\ref{planchar} to be
5.6~g\,cm$^{-3}$. This is comparable to that of Mercury and Io
although the reflective properties at the surface are independent of
the composition. Additionally, the similar bulk density to Mercury
implies a greater concentration of volatiles resulting in high
densities at the core with relatively low-density near the surface.


\subsection{Variable Incident Stellar Flux}
\label{varflux}

The geometric albedo can have a spatial dependence which is due to a
variable amount of stellar flux being received at various points on
the planetary surface. Consider the cases of Mercury and
Venus. Mercury experiences a large range of day-side surface
temperatures due to a steep temperature gradient between the equator
and the poles. The 3:2 spin-orbit resonance of Mercury results in the
periodic cooling off of the surface as it crosses the terminator into
the night-side of the planet. In the case of Venus, the slow
retrograde rotation would also allow for extreme temperature gradients
both on the day-side and at the day/night boundary if it were not for
the thick atmosphere which is exceptionally efficient at
redistributing the trapped thermal radiation.

From the planetary radius and semi-major axis for 55~Cnc~e shown in
Table~\ref{planchar}, the stellar flux at the orbital distance of the
planetary poles is $\sim 10$\% less than the flux at the
equator. Spreading this flux onto the surface of the planet at a given
latitude then yields a surface flux which is further reduced by the
cosine of the angle the host star is from zenith at that location.
However, we assume that the planet is tidally locked such that no
cooling of the day-side surface ever occurs. For the various surface
models described in Section \ref{phase} we then consider a constant
albedo for the day-side surface. For the molten surface model in
particular, the variable stellar flux is sufficient at all latitude to
result in the necessary sustained temperatures for melting the surface
silicate materials. This is described further in Section \ref{phase}.


\section{Phase Variations}
\label{phase}

\begin{figure*}
  \begin{center}
    \begin{tabular}{cc}
      \includegraphics[angle=270,width=8.2cm]{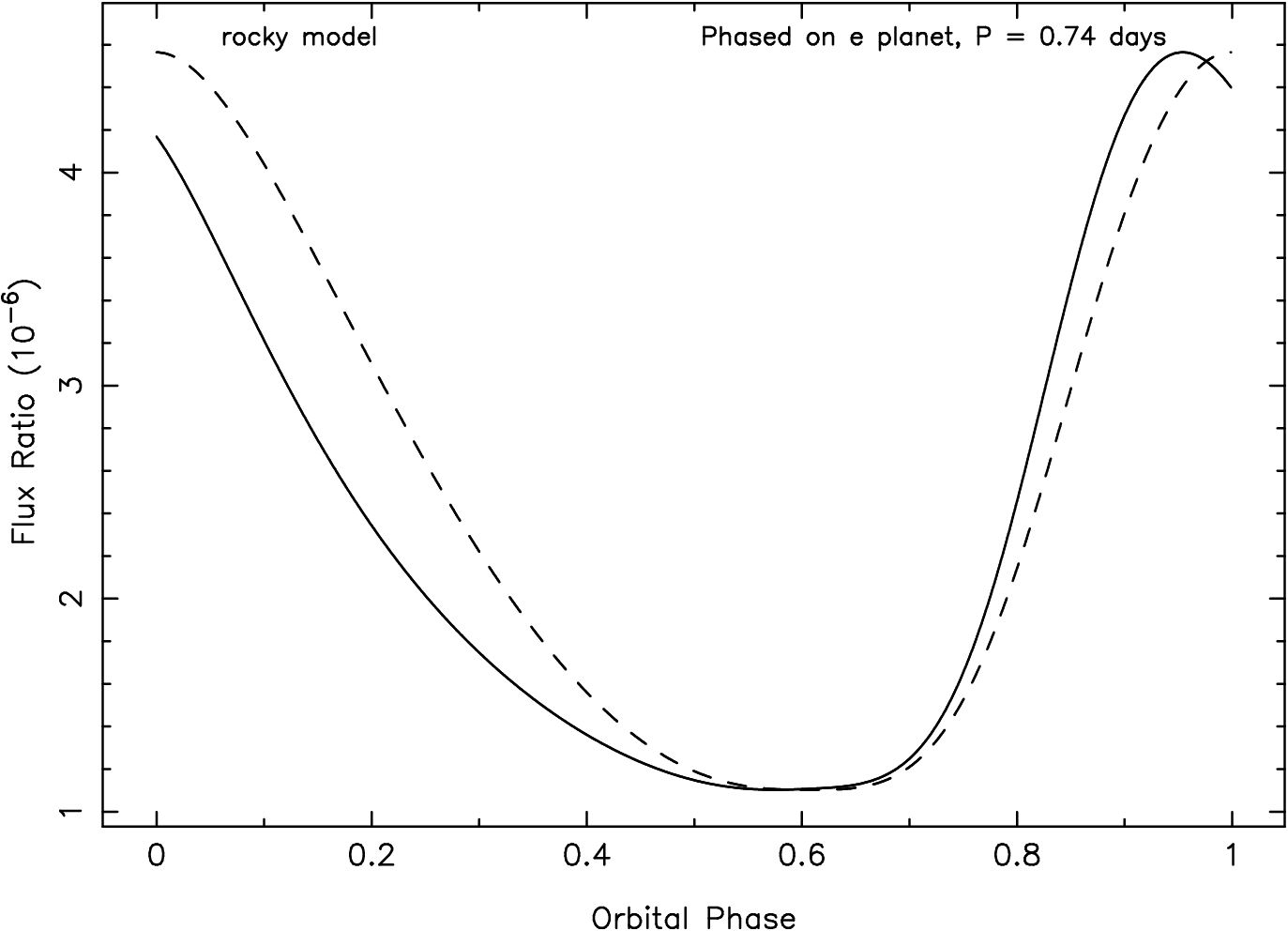} &
      \includegraphics[angle=270,width=8.2cm]{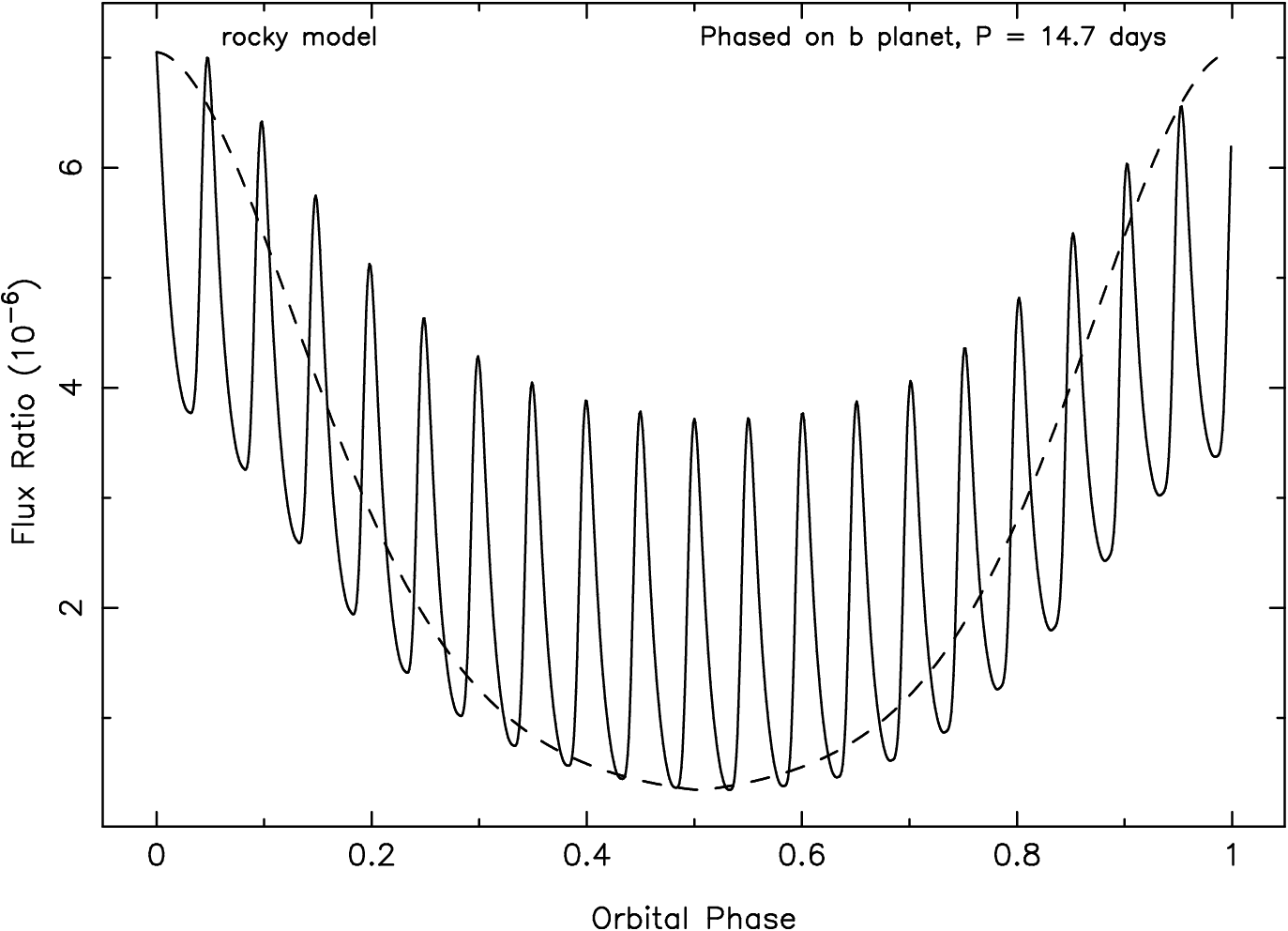} \\
      \includegraphics[angle=270,width=8.2cm]{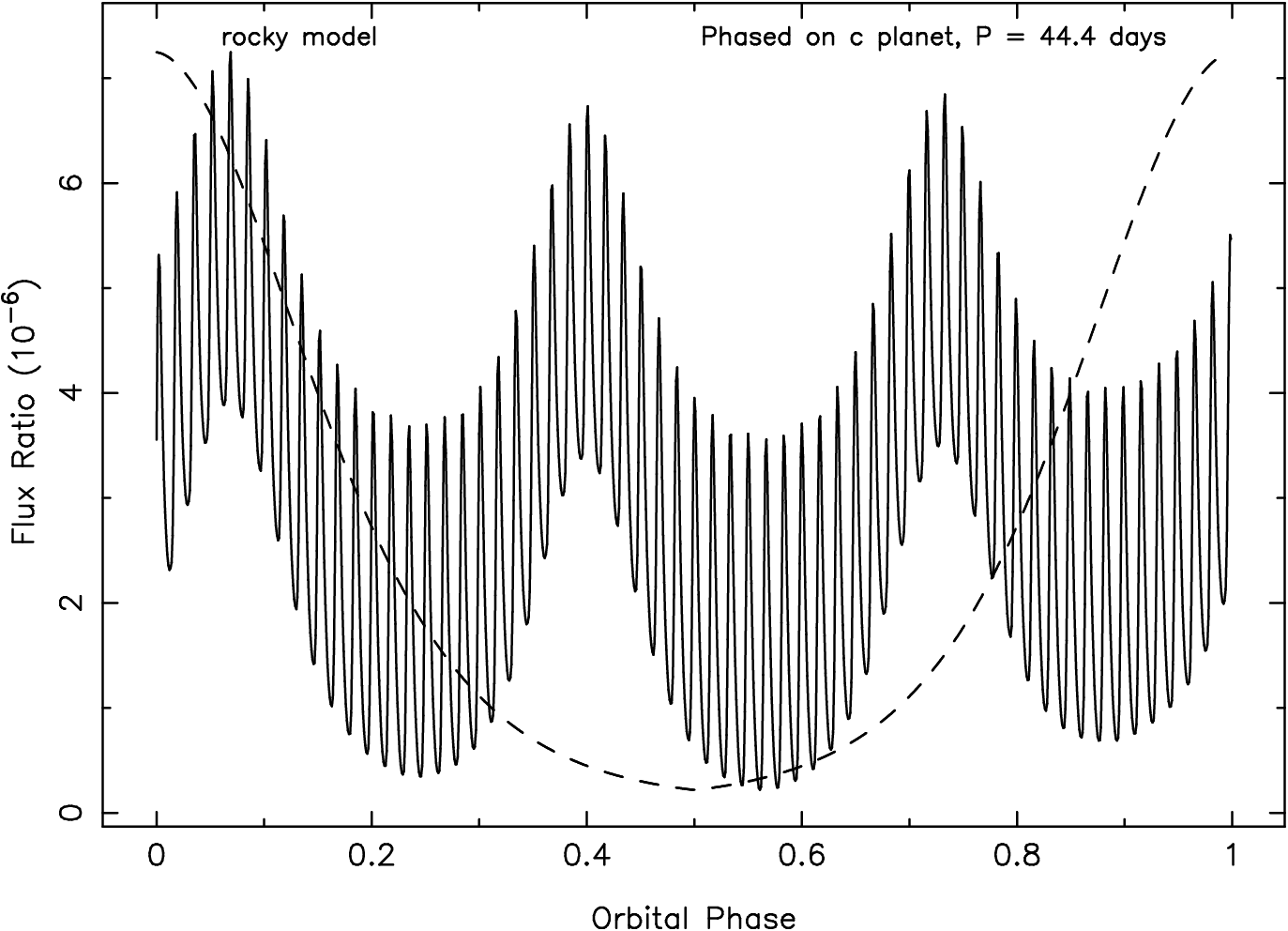} &
      \includegraphics[angle=270,width=8.2cm]{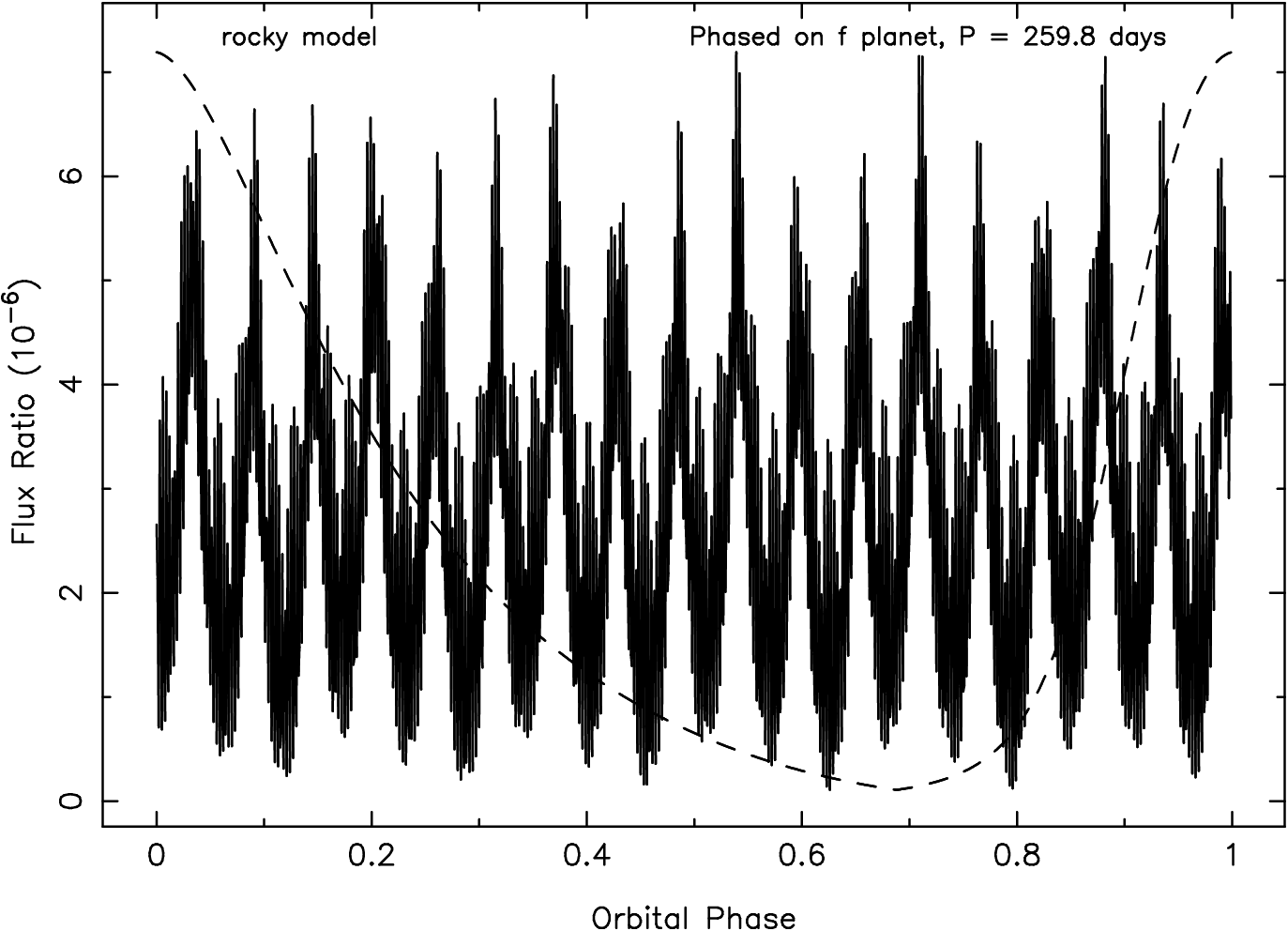} \\
    \end{tabular}
  \end{center}
  \caption{The flux ratios (solid line) of the five 55 Cnc system
    planets to the host star, where the surface of the e planet is
    assumed to be rocky with a geometric albedo of 0.1. The time scale
    in each plot has been set to show one complete orbital phase of
    planets e (top-left), b (top-right), c (bottom-left), and f
    (bottom-right). In each plot, the dashed line indicates the
    normalized phase function for the planet on which the figure is
    phased (see Section \ref{rockysection}). These plots show that the
    combined phase signatures of all five planets are dominated by the
    two inner-most planets (top-left), beyond which little information
    can be extracted for the others without exquisite photometric
    precision.}
  \label{rocky}
\end{figure*}

Here we simulate the phase variations of the 55 Cnc system. The phase
algorithms are based upon the formalism of \citet{kan10}. For each
planet, the phase angle $\alpha$ is defined to be zero at superior
conjunction and is described by
\begin{equation}
  \cos \alpha = \sin (\omega + f) \sin i
  \label{phaseangle}
\end{equation}
where $f$ is the true anomaly and $i$ is the inclination of the
orbit. The phase function of a Lambert sphere assumes isotropic
scattering of incident flux over $2 \pi$ steradians and is described
by
\begin{equation}
  g(\alpha,\lambda) = \frac{\sin \alpha + (\pi - \alpha) \cos
    \alpha}{\pi}
\end{equation}
and is used for the rocky and molten surface models described
below. For the atmosphere model, we adopt the empirically derived
phase function of \citet{hil92} which is based upon observations of
Jupiter and Venus. This approach contains a correction to the
planetary visual magnitude of the form
\begin{equation}
  \Delta m (\alpha) = 0.09 (\alpha/100\degr) + 2.39
  (\alpha/100\degr)^2 -0.65 (\alpha/100\degr)^3
\end{equation}
leading to a phase function given by
\begin{equation}
  g(\alpha) = 10^{-0.4 \Delta m (\alpha)}
\end{equation}
which we refer to as the Hilton phase function and allows for
non-isotropic (cloud) scattering. Here we confine our study to optical
wavelengths centered on 550~nm. This places the study near the peak
response of the Kepler and MOST detectors. It is possible for there to
be a thermal component to the phase variation at these wavelengths,
such as that predicted by the models of \citet{dem11c} for
Kepler-7b. In addition, \citet{wel10} speculate that the Kepler
observations of HAT-P-7b may include a thermal component due to the
planet not emitting as a blackbody as is often assumed. Depending upon
the specific surface scattering properties, this may lead to a
non-negligible underestimate of the phase amplitude for the e planet
that will be interesting to resolve when more precision photometry is
acquired. Note that studies of flux ratio dependencies on wavelength
have been undertaken by \citet{sud05}.


\subsection{Rocky Surface Model}
\label{rockysection}

Given the size, mass, and density of the planet e, as well as its
extreme proximity to the host star, it is likely that any atmosphere
that may once have existed has since evaporated and been stripped away
by the stellar irradiation. We thus first consider the case that the
surface of the planet is of a rocky form, similar to the surface of
Mercury. The density of the planet implies a different composition to
Mercury (indeed to all the solar terrestrial planets), but the heavier
materials likely reside closer to the center of the planet than the
surface.

Shown in Figure \ref{rocky} are the expected phase signatures of the
55~Cnc system using the rocky surface model for the planet e, where
the flux ratio refers to the combined flux of all the planets to that
of the host star. Hence the phase variation effects for all planets
are included in each panel, but panels are zoomed-in on the orbital
phase of the e, b, c, and f planets respectively and are thus phased
on those particular planets in each case. In all four panels, the
dashed line corresponds to the phase function of the planet on which
the figure is phased. The phase function as shown has been normalized
to the y-scale of each plot to show the time-dependent contribution of
the planet to the total phase curve.  For the purposes of this
simulation, we first assume that all planets are located at periastron
passage, then we move time forward to start the phase curve where the
outer-most planet in each plot is located at a phase angle of
zero. This can be seen in the phase function shown in each panel which
is at maximum value at an orbital phase of zero. Therefore these
simulations represent a specific orbital configuration, the effect of
which we discuss further in Section \ref{config}.

\begin{figure*}
  \begin{center}
    \begin{tabular}{cc}
      \includegraphics[angle=270,width=8.2cm]{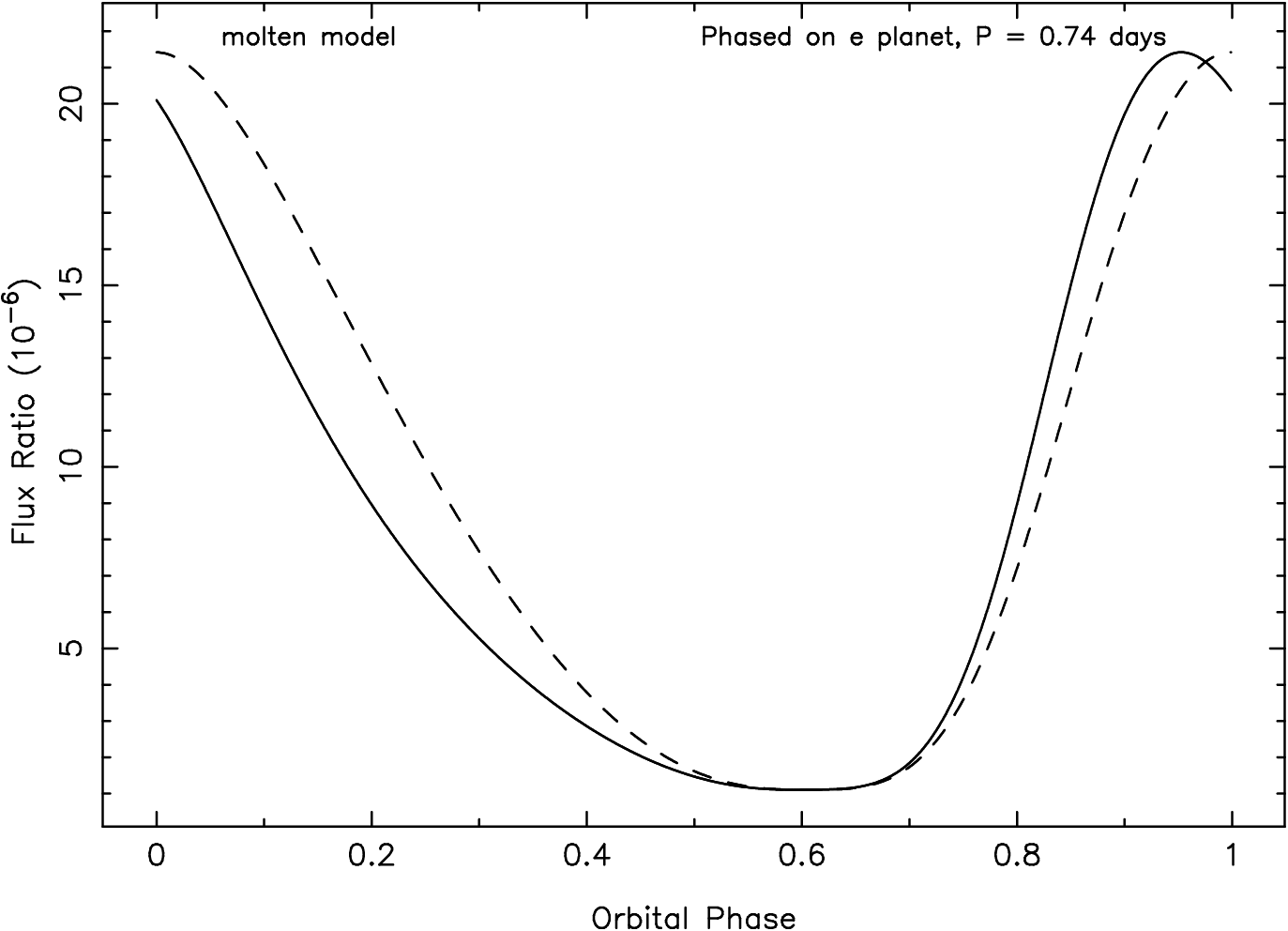} &
      \includegraphics[angle=270,width=8.2cm]{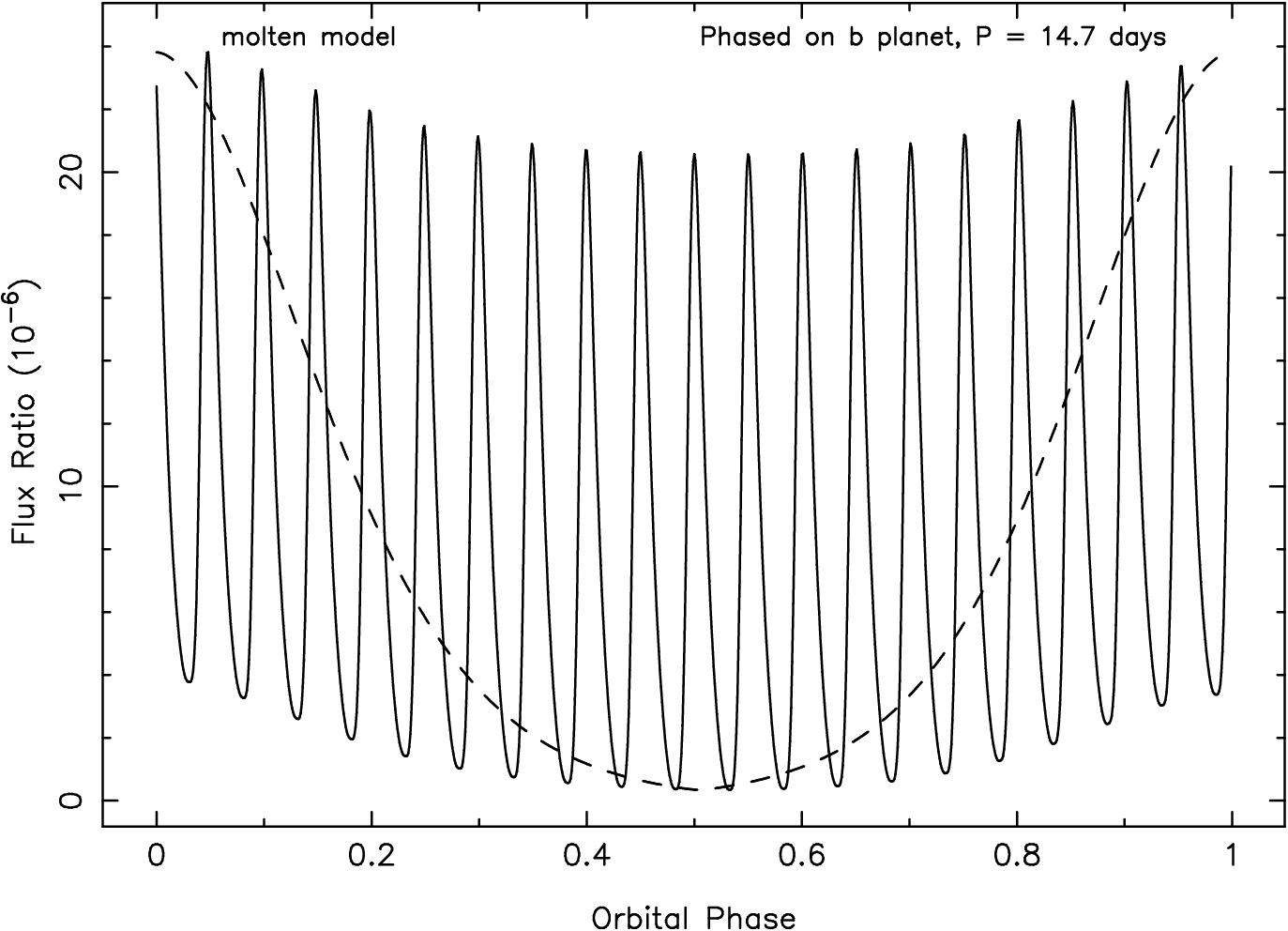}
    \end{tabular}
  \end{center}
  \caption{As for the top two panels of Figure \ref{rocky}, but the e
    planet is now assumed to have a molten surface with a geometric
    albedo of 0.6 (see Section \ref{moltensection}).}
  \label{molten}
\end{figure*}

The phase curves in Figure \ref{rocky} show that the total phase
variation of the system is dominated by planets e and b whose
independent flux ratio amplitudes are almost identical; $3.4 \times
10^{-6}$ and $3.3 \times 10^{-6}$ respectively. The flux ratio
amplitude of the c planet is almost an order of magnitude less; $7.0
\times 10^{-7}$. Note the asymmetric modulation for the e and f
planets (top-left and bottom-right panels) due to their respective
eccentric orbits. The maximum flux ratio amplitude of the f planet is
$1.1 \times 10^{-7}$ and occurs near the periastron passage near phase
0.7. The outer-most planet, d, is sufficiently far away from the host
star that it contributes a negligible amount of flux to the total flux
ratio; $4.0 \times 10^{-9}$. These results are summarized in Table
\ref{planchar}.


\subsection{Molten Surface Model}
\label{moltensection}

\citet{win11} estimate a surface temperature of 2800~K at the
substellar point if the planet is tidally-locked. Even if the heat is
somehow redistributed, the equilibrium temperature will still be as
high as 1980~K. This is well above the melting point for igneous
rock. In addition, the best-fit solution by \citet{daw10} contains a
non-zero eccentricity for planet e which would create a ``super-Io''
effect, as described by \citet{bar10}. We thus here consider the
entirely plausible case of a molten surface for the inner planet.

Figure \ref{molten} is equivalent to the top two panels of Figure
\ref{rocky} in that they are zoomed-in to the phases of planets e and
b. The much higher albedo causes the e planet to become dominant in
the phase curve shown in the right panel (note the different ordinate
scales). The e planet now has a flux ratio amplitude of $2.0 \times
10^{-5}$ which is almost an order of magnitude higher than that for
the b planet.


\subsection{Atmosphere Model}
\label{atmossection}

The e planet is unlikely to harbor an atmosphere under the extreme
conditions of its environment \citep{win11}. Here we consider this
possibility for completeness and as a direct comparison with the other
two presented scenarios for the surface of planet e. In this case we
use the Hilton rather than the Lambert phase function to represent the
non-isotropic scattering of the atmosphere. The non-uniform incident
stellar flux described in Section \ref{varflux} is insufficient to
allow an non-uniform albedo since the reflective consendates will be
removed from the upper atmosphere regardless of latitude under such
extreme temperatures. As shown in Table \ref{planchar}, the mean
geometric albedo for the atmosphere model is only slightly higher than
that for the rocky surface due to these reflective condensates being
effectively removed.

\begin{figure*}
  \begin{center}
    \begin{tabular}{cc}
      \includegraphics[angle=270,width=8.2cm]{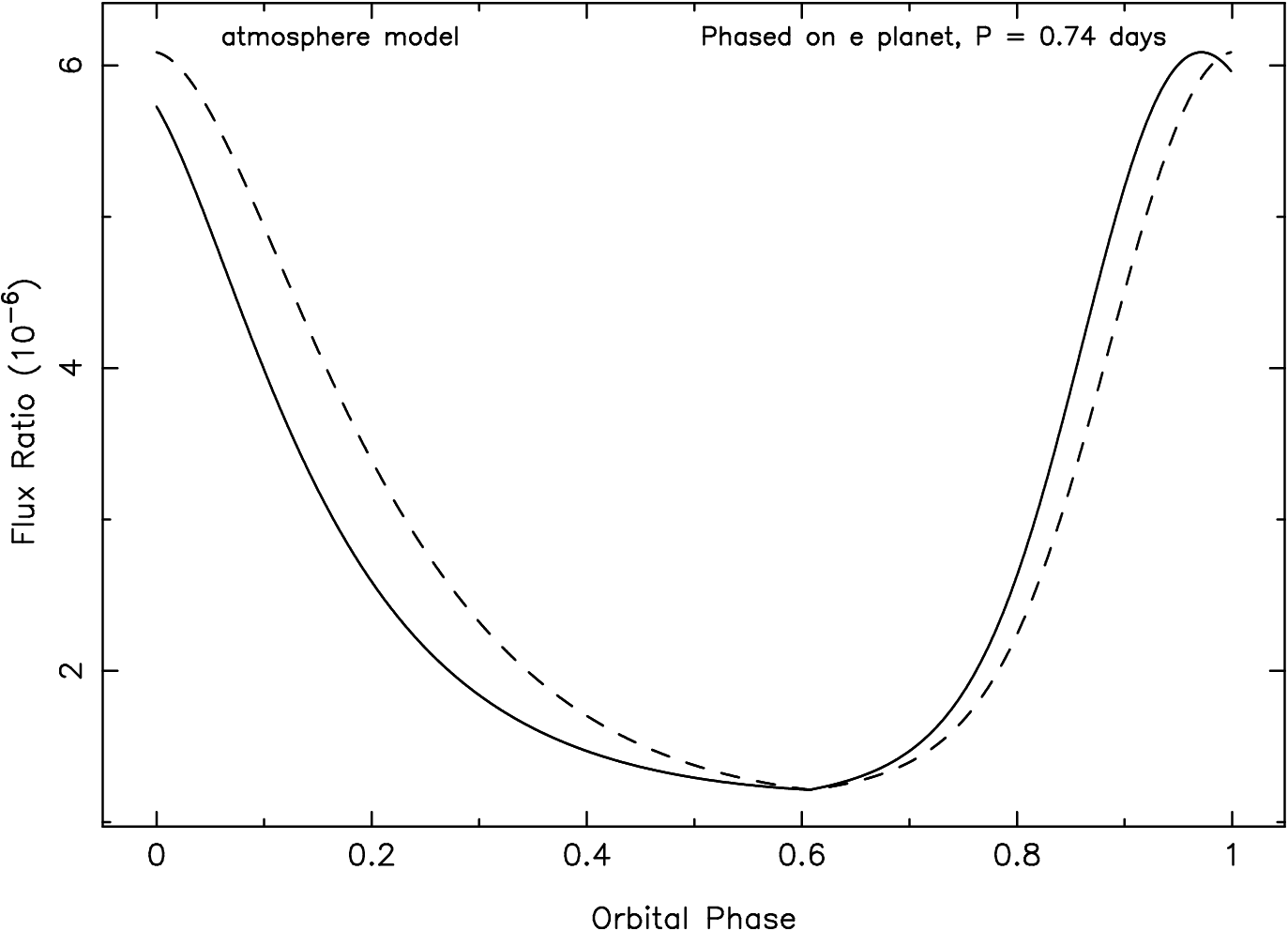} &
      \includegraphics[angle=270,width=8.2cm]{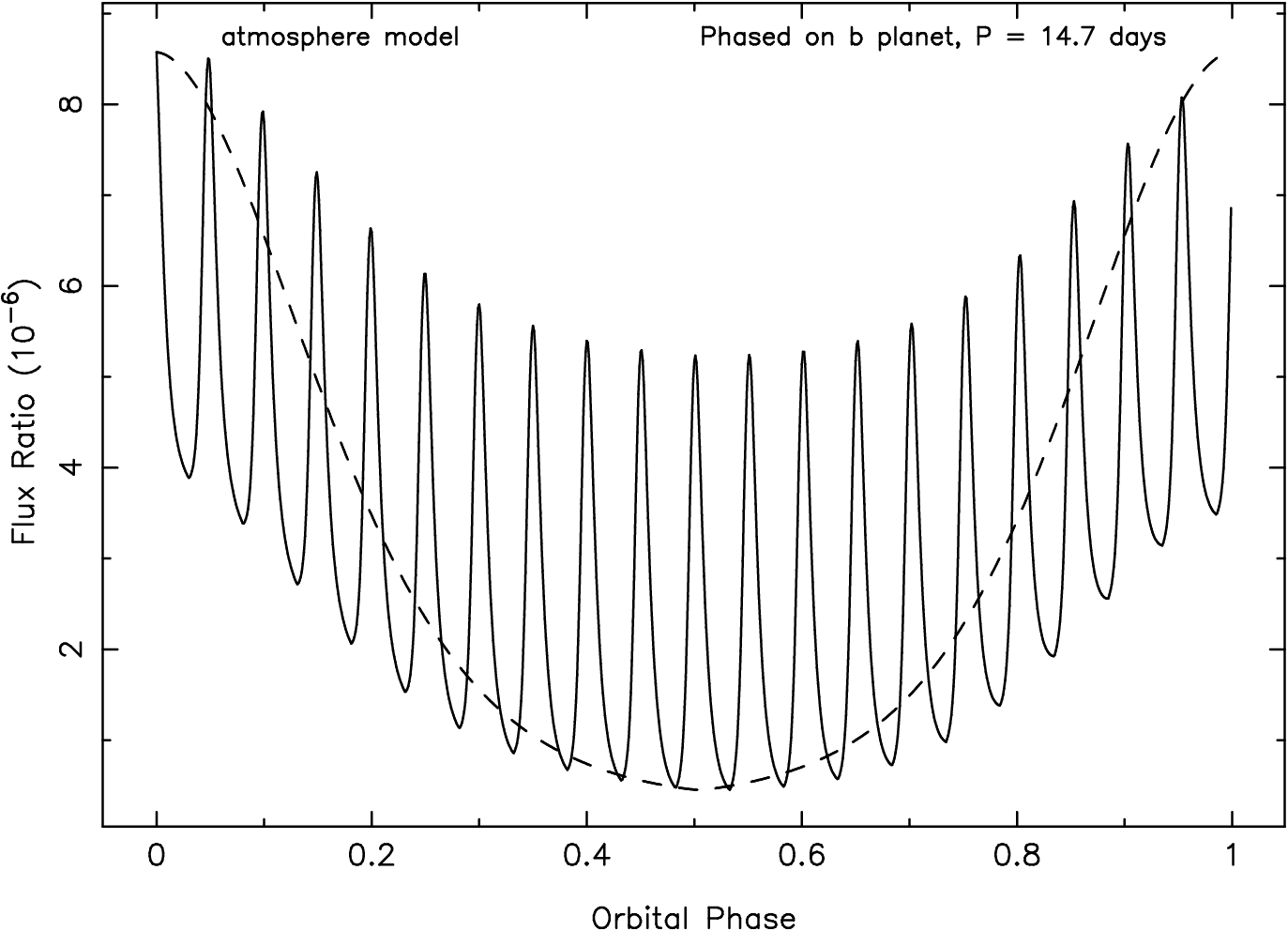}
    \end{tabular}
  \end{center}
  \caption{As for the top two panels of Figure \ref{rocky}, but the e
    planet is now assumed to have an atmosphere with non-isotropic
    scattering properties (see Section \ref{atmossection}).}
  \label{atmos}
\end{figure*}

In Figure \ref{atmos} we see the combined calculated flux ratio zoomed
to the phases of the e and b planets where, once again, the ordinate
scales have increased relative to Figure \ref{rocky}. This model
produces a flux ratio amplitude for planet e of $4.9 \times 10^{-6}$
which slightly exceeds the amplitude expected from the b planet. We
discuss being able to distinguish between this model and the rocky
surface model in Section \ref{signal}.


\subsection{Orbital Configurations and Phase Modulation}
\label{config}

The simulation results provided above describe a specific starting
configuration for the planets which is based upon their periastron
passages and phase angles. However, for observations at some random
epoch, this configuration will be arbitrary in nature and the phase
curves shown here will not necessarily match that which is
observed. In particular, the relative phases of the two planets which
contribute the bulk of the total reflected light, e and b, will result
in phase modulations. Thus there will be optimal orbital
configurations for which to search for the phase signatures. The
maximum amplitude of the flux ratio will approximately occur where
both of the planets are simultaneously close to zero phase
angle. However, the extraction of only the phase information for
planet e will be best achieved when planet b is near inferior
conjunction. Fortunately, the orbital periods of these two planets are
relatively short which allows the scheduling of such optimal
observating times to be straightforward.

If one requires both high signal-to-noise throughout the measurement,
and folding of multiple orbits of planet e, then one needs to be
mindful of the contributions of the outer planets as they progress in
their orbits. For each orbit of planet e, the outer planets will shift
in phase by 0.05, 0.017, 0.0028, and 0.00014 respectively. Through a
single orbit this will have minor effects on the total phase
variation, but over multiple orbits, will begin to show modulation
effects which will dampen the signature of planet e, thus resulting in
an incorrect estimate of the flux ratio amplitude. Deriving accurate
ephemerides through further RV measurements will help to curtail such
effects in photometric monitoring of the phase variations.


\section{Signal Detectability}
\label{signal}

Here we assess the detectability of the planetary flux ratios and the
possible effects of stellar variability.


\subsection{Instrumentation Requirements}

From the expected phase amplitude of the two inner-most planets of
55~Cnc, the instrumentation requirement for successful detection of
the phase signatures is photometry with a relative photometric
precision of at least $\sim 10^{-6}$. For the e and b planetary phase
variations, long-term stability of high-precision photometry is not
required. The Kepler mission is already achieving this precision for
significantly fainter stars, though it should be noted that the
exquisite photometers of Kepler were designed to perform such a
task. An example of the high-standard of Kepler precision is the
detection of phase and ellipsoidal variations by \citet{wel10} using
only the Q1 Kepler data, where the amplitude of this variation is $3.7
\times 10^{-5}$. Using the STIS instrument on the Hubble Space
Telescope (HST), \citet{bro01} obtained a precision of $1.1 \times
10^{-4}$ per 60 second integration observation during primary transits
of HD~209458b. Such high time resolution is not required for phase
variation observations and so binning these data would improve the rms
scatter. A future mission that would allow such observations of 55~Cnc
to be carried out include the James Webb Space Telescope (JWST) with
the NIRCam instrument, though wavelength range of this instrument
would include a substantial thermal component of the phase variation
which would need to be accounted for. The minimum precision mentioned
above would adequately confirm or rule out a molten surface for planet
e. Further RV characterization of the orbits for the 55~Cnc planets
will allow one to accurately predict both the amplitude of the
predicted phase signature and times of maximum and minimum flux
ratios. This knowledge will help to distinguish the phase signatures
from instrumental drift effects.

Future possabilities also exist from the ground, although one needs to
also contend with the offsets from night-to-night variations. There
are several large telescopes under development that are capable of
meeting the challange of very high photometric precision, such as the
European Extremely Large Telescope (E-ELT), the Thirty Meter Telescope
(TMT), and the Giant Magellan Telescope (GMT). It has also been
demonstrated by \citet{col10} that precision photometry of $< 0.05$\%
can be achieved with the 10.4m Gran Telescopio Canarias through the
use of narrow-band filters. These was conducted for observing the
signatures of known transiting planets which is possible to achieve
within a single night. Longer term monitoring of phase signatures will
require careful accounting for the aforementioned nightly variations
in addition to the airmass corrections throughout a night.


\subsection{Stellar Variability of 55 Cnc}

If one is able to accomplish the required level of photometric
precision, the greatest impediment to studying the planetary phase
variations will be the intrinsic stellar variability. An analysis of
Kepler data by \citet{cia11} found that most dwarf stars are stable
down to the the precision of the Kepler spacecraft, with G-dwarfs
being the most stable of the studied spectral types. The main cause of
photometric variability in F--G--K stars is starspots and
rotation. The rotation period for 55~Cnc has been measured on numerous
occasions through photometric variations. \citet{sim10} calculate a
rotation period of 44.1~days and \citet{fis08} measure a rotation
period of 44.7~days. \citet{win11} also observed variation of the
order 10$^{-4}$ which is assumed to be due to both stellar activity
and rotation. For the c planet where the orbital period is close to
the rotation period of the star, the variation due to phase and
rotation may be difficult to disentangle. The peaks in the power
spectrum from a fourier analysis of the photometry may separate to a
degree where the starspot variability can be isolated from the phase
signature. The known phase of the planet from the RV analysis will be
the greatest aid in discriminating these two signals. It should also
be noted that there is an M dwarf binary companion, 55~Cnc~B, with an
angular separation of 84.7\arcsec (1150~AU) \citep{mug06}, so it is
unlikely to be inside a photometric aperture.


\section{Conclusions}

The multiple-planet system of 55~Cnc presents many opportunities to
understand and characterize a system around a solar analog with very
different characteristics to our own system. The discovery that the
inner-planet transits enhances the opportunities since we can place
greater constraints on experiments designs to investigate these
properties. Here we have specifically addressed the method of
detecting the phase variations of the planets. These results show that
the inner two planets have flux ratio amplitudes which are comparable
to what has already been detected by Kepler around much fainter stars
and will be accessible to next-generation ground and space-based
observing platforms if not before. Which of the two planets dominates
the phase signature depends upon if the planet has a rocky or molten
surface, with the molten surface model producing a dominance of planet
e to the signature due to the higher reflective properties of the
surface. The phase variation of the e planet for a rocky surface is
almost indistinguishable from that of an atmosphere model at the level
of the precision requirements for detection, 10$^{-6}$. These two
models have very similar surface albedos and it makes little
difference whether one assumes isotropic or non-isotropic (cloud)
scattering at such small star--planet separations. The outer planets
do not present a significant impediment for the detection of the e and
b phase variations but one needs to be aware of the expected phase
modulations if multiple phase observations are undertaken to boost
signal-to-noise. We have not considered here the thermal component of
the flux from the e planet, whose variation could be significant if
indeed the planet is tidally-lock as suggested by \citet{win11}.

Even though the transit depth is relatively small, the high
signal-to-noise possible from transiting planets such as 55~Cnc~e
demonstrate the value of such objects. The best hope then for more
such transiting planets around bright stars lies through investigation
of known exoplanets discovered using the RV technique
\citep{kan09}. Projects such as the Transit Ephemeris Refinement and
Monitoring Survey (TERMS) seek to accomplish just that which will
hopefully lead to further characterization opportunities.


\section*{Acknowledgements}

The authors would like to thank Alan Boss for insightful discussions
as well as Brice-Olivier Demory and Joshua Winn for their useful
inputs. We would also like to thank the anonymous referee, whose
comments greatly improved the quality of the paper.


\end{document}